\documentclass [12pt] {article}
\usepackage{amsfonts}
\usepackage{amssymb}

\newcommand{\qed} {\hspace {0.1in} \rule {1.5mm} {3.5mm}}

\newtheorem{lemma}{Lemma}[section]

\newtheorem{theorem}{Theorem}

\newtheorem{conjecture}{Conjecture}
\newtheorem{proposition}{Proposition}[section]

\def\limi{\lim_{i\to\infty}}
\def\limn{\lim_{n\to\infty}}

\def\e{\epsilon}

\def\deg{\mbox{deg}\,}
\def\ke{\mbox{Ker}\,}
\def\im{\mbox{Im}\,}
\def\rank{\mbox{Rank}\,}

\def\mat{\mbox{Mat}\,}

\def\dim{{\rm dim}}

\def\Fo{F$\mbox{\o}$lner}
\def\<{\langle}
\def\>{\rangle}

\def\proof{\smallskip\noindent{\it Proof.} }

\def\bR{{\mathbb R}}

\def\deg{\mbox{deg}\,}

\def\cA{\mbox{$\cal A$}}

\def\cR{\mbox{$\cal R$}}

\def\cR{\mbox{$\cal R$}}

\def\lan{\langle}

\def\ran{\rangle}

\def\to{\rightarrow}

\def\ws{\widetilde{\sigma}}
\begin{document}
\title{Aperiodic order, integrated density of states and
the continuous algebras of John von Neumann}
\author{\sc G\'abor Elek
\footnote {The Alfred Renyi Mathematical Institute of
the Hungarian Academy of Sciences, P.O. Box 127, H-1364 Budapest, Hungary.
email:elek@renyi.hu, Supported by OTKA Grants T 049841 and T 037846}}
\date{}
\maketitle
\vskip 0.2in
\noindent{\bf Abstract.}  In \cite{LS2} Lenz and Stollmann proved
the existence of the integrated density of states in the sense
of uniform convergence of the distributions for certain operators
with aperiodic order. The goal of this paper is to establish
a relation between aperiodic order, uniform spectral convergence and
the continuous algebras invented by John von Neumann. We illustrate
the technique by proving the
uniform spectral convergence for random Schr\"odinger operators on
lattices with finite site probabilities,
percolation Hamiltonians and for the pattern-invariant
operators of self-similar graphs.

\vskip 0.3in
\newpage
\section{Introduction}
First, let us recall the notion of uniform spectral convergence studied
by Lenz and Stollmann \cite{LS1}, \cite{LS2}. Let $G$ be a simple, connected
infinite graph with bounded vertex degrees, with vertex set $V$ and
edge set $E$. Let $\Delta_G:L^2(V)\to L^2(V)$ be the combinatorial
Laplacian operator:
$$\Delta_G\,f(x)=\deg(x)f(x)-\sum_{(x,y)\in E}f(y)\,.$$
A \Fo -sequence in $G$ is a sequence of finite subsets
$\{Q_n\}^\infty_{n=1}\subset V$ such that $\frac{|\partial Q_n|}{|Q_n|}\to 0$,
where
$$\partial Q_n:=\{p\in Q_n\,\mid \mbox{there exists}\,q\notin Q_n, (p,q)\in
E\}\,.$$
For a subset $A\subset V$, let $\Delta_A:L^2(A)\to L^2(A)$ be
the Laplacian on the subgraph $G_A$ spanned by the vertices of $A$.
If $A$ is finite, then it is a positive, self-adjoint linear transformation.
The normalized spectral distribution function of $\Delta_A$ is defined
the following way:
$N_{\Delta_A}(\lambda):=\{$ the number of eigenvalues of $A$ counted with
multiplicities, not greater than $\lambda$, divided by $|A|\}$.
We say that the integrated density of states exists in the sense
of uniform convergence of the distributions (shortly: the uniform
spectral convergence exists) if there is a monotone function
$N_{\Delta_G}$ such that for any \Fo -sequence $\{Q_n\}^\infty_{n=1}$
the functions $N_{\Delta_{Q_n}}$ uniformly converge to $N_{\Delta_G}$.
It is important to note that weak spectral convergence, that is
$$\limn N_{\Delta_{Q_n}}(\lambda)= N_{\Delta_G}$$
for all points of continuity of $N_{\Delta_G}$ is generally much easier
to prove (see the discussion in Section \ref{last}).
Lenz and Stollmann proved the existence of universal spectral
convergence in the case of aperiodic order for Laplacian-type operators
defined by nice Delone-sets of $\bR^d$. In our paper we study the
same problem form a ring theoretical point of view.
We introduce the spectral functions for the elements of continuous
algebras defined by Goodearl \cite{Good} as a generalization of
the original construction of John von Neumann \cite{Neu}.
These rings $\cR$ are complete metric rings equipped with rank functions.
We shall see in Proposition \ref{spec} that if
$\{T_n\}_{n=1}^\infty \subset \cR$ converge to $T\in \cR$, then the
spectral functions of $T_n$ converge uniformly to the spectral
function of $T$. We shall use this fact to prove uniform spectral
convergence for random and deterministic operators.
The examples considered in the paper are:
\begin{itemize}
\item Random Schr\"odinger operators on lattices with finite
site probabilities.
\item Percolation Hamiltonians.
\item Pattern-invariant operators on self-similar graphs.
\end{itemize}

\noindent
{\bf Acknowledgement}: I would like to thank the organizers of the
MFO-Mini-Workshop on ``$L^2$-invariants and the Integrated Density of
States'' for inviting me.
\section{The Goodearl-von Neumann construction} \label{sec2}

In this section we recall Goodearl's construction of rank rings
\cite{Good} via Bratteli diagrams. This is a generalization of
the original construction of John von Neumann \cite{Neu}.
Let $\cR_1, \cR_2,\dots$ be finite dimensional algebras over the
real numbers,
$$\cR_i:=\bigoplus^{k_i}_{\alpha=1} \mat_{n_{i,\alpha}\times n_{i,\alpha}}
(\bR)\,.$$
Next we consider injective unital algebra homomorphisms 
$\phi_i:\cR_i\to \cR_{i+1}$.
For each $i\geq 1$ we have $k_i$ {\it nodes}
 $\{a^i_{\alpha}\}^{k_i}_{\alpha=1}\,.$ Also, for each pair
$(a^i_\alpha, a^{i+1}_\beta)$, $1\leq \alpha \leq\ k_i$, $1\leq \beta 
\leq k_{i+1}$
we have {\it weights} $w_{\alpha,\beta}$ satisfying the following
compatibility condition:
\begin{itemize}\item
For any $1\leq \beta \leq k_{i+1}$:
$$n_{i+1,\beta}=\sum^{k_i}_{\alpha=1} w_{\alpha,\beta} n_{i,\alpha}\,.$$
Thus this system defines diagonal imbeddings $\phi_i:\cR_i\to\cR_{i+1}$.
\end{itemize}
One can associate Markov-transition probabilities \\ to the weight system:
$M(\beta,\alpha):=\frac{w_{\alpha,\beta} n_{i,\alpha}}{n_{i+1,\beta}}\,.$
A rank function on the inductive limit is defined as follows.
Let $p_{i,\alpha}>0$ be real numbers satisfying the relations:
\begin{equation} \label{e21}
\sum^{k_i}_{\alpha=1} p_{i,\alpha}=1\,
\end{equation}
\begin{equation} \label{e22}
\mbox{For any}\,\, 1\leq \alpha \leq k_i \,\,\,\, p_{i,\alpha}=
\sum^{k_{i+1}}_{\beta=1}
M(\beta,\alpha) p_{i+1,\beta}\,.
\end{equation}
Then for each $i\geq 1$ one has a normalized rank function
$r_i:\cR_i\to\bR$, defined as
$$r_i(A_{i,1}\oplus A_{i,2}\oplus\dots\oplus A_{i,k_i}):=
\sum^{k_i}_{\alpha=1} p_{i,\alpha}\frac {\rank
  A_{i,\alpha}}{n_{i,\alpha}}\,.$$
Then if $A=(A_{i,1}\oplus A_{i,2}\oplus\dots\oplus A_{i,k_i})\in\cR_i,
\,r_i(A)=r_{i+1}(\phi_{i+1}(A))\,.$ Hence we defined a
normalized rank function $r$ on the inductive limit
$\lim_{\stackrel{\phi_i}{\to}} \cR_i$.
Clearly the rank $r$ satisfies the following conditions:
\begin{itemize}
\item $r(A)\geq 0$ for any $A\in \lim_{\stackrel{\phi_i}{\to}} \cR_i$.
\item $r(1)=1, r(0)=0$ and $r(A)=0$ if and only if $A=0$.
\item $r(A+B) \leq r(A)+r(B)\,.$
\item $r(AB)\leq r(A),\, r(AB)\leq r(B)\,.$
\end{itemize}
Therefore  $d(A,B):=r(A-B)$ defines a metric on the inductive limit ring.
Its completion is the continuous regular ring $\cR$ and the
rank $r$ extends to $\cR$ satisfying the four conditions above.
\section{Spectral functions}
Let us consider the inductive limit system
$$\cR_1\stackrel{\phi_1}{\to} \cR_2 \stackrel{\phi_2}{\to}\dots$$
as in Section \ref{sec2}. For each $i\geq 1$ consider the vectorspace
$\bigoplus^{k_i}_{\alpha=1} \bR^{n_{i,\alpha}}=V_i$. The algebra
$\cR_i=\bigoplus^{k_i}_{\alpha=1} \mat_{n_{i,\alpha}\times n_{i,\alpha}}
(\bR)$ acts on $V_i$ the natural way.
Now we define a dimension function on the $\cR_i$-submodules of $V_i$.
Note that any such $\cR_i$-submodule $L$ is in the form
$\bigoplus^{k_i}_{\alpha=1} L_{i,\alpha}$,
where $L_{i,\alpha}\subseteq \bR^{n_{i,\alpha}}$ is a linear subspace. Then
$$\dim_{\cR_i}(\bigoplus^{k_i}_{\alpha=1} L_{i,\alpha}):=
\sum^{k_i}_{\alpha=1} p_{i,\alpha}\frac{\dim_{\bR} L_{i,\alpha}}{n_{i,\alpha}}
\,.$$ Then of course, $\dim_{\cR_i}(V_i)=1$.
Let $T\in \cR_i$, then we call a $\cR_i$-submodule $L\subseteq V_i$ a
$\lambda^-$-space of $T$ if $\|T(v)\|\leq \lambda \|v\|$, for any
$v\in L$, where $\|v\|$ is the Euclidean norm of $v$.
Similarly, we call a $\cR_i$-submodule $N\subseteq V_i$ a
$\lambda^+$-space of $T$ if $\|T(v)\| >\lambda \|v\|$, for any
$0\neq v\in N$.
Now we define $\sigma^i_T(\lambda)$ as the maximal dimension
(with respect to $\dim_{\cR_i}$ !) of all the $\lambda^-$-spaces of $T$
and $\ws^i_T(\lambda)$ as the maximal dimension
 of all the $\lambda^+$-spaces of $T$.
\begin{lemma} \label{l1} For any $T\in \cR_i$,
$\sigma^i_T(\lambda)+\ws^i_T(\lambda)=1$. If
 $T=\bigoplus^{k_i}_{\alpha=1}T_{\alpha}$ is a positive, self-adjoint element
then $\sigma^i_T(\lambda)$ equals to $\sum^{k_i}_{\alpha=1} p_{i,\alpha}
N_{T_{\alpha}}(\lambda)$,
where $N_{T_{\alpha}}$ is the normalized spectral distribution
function, defined in the Introduction.
\end{lemma}
\proof
Let $S\in \cR_i$ be the positive self-adjoint matrix such that
$S^2=T^*T$. Then $\sigma^i_S(\lambda)= \sigma^i_T(\lambda)$
and $\ws^i_S(\lambda)= \ws^i_T(\lambda)$. Indeed
$$\|T(v)\|\leq \lambda \|v\| \Leftrightarrow
\lan T(v),T(v)\ran\leq \lambda^2 \|v\|^2 \Leftrightarrow
\lan T^*T(v),v \ran\leq \lambda^2 \|v\|^2 \Leftrightarrow $$ $$ \Leftrightarrow
\lan S^2(v),v\ran \leq \lambda^2\|v\|^2 \Leftrightarrow
\|S(v)\|\leq \lambda\|v\|\,.$$
Hence in order to prove the equality  $\sigma^i_T(\lambda)+\ws^i_T(\lambda)=1$
it is enough to suppose that $T$ is positive self-adjoint. Clearly,
$\sigma^i_T(\lambda)+\ws^i_T(\lambda)\leq 1$
since the intersection of a $\lambda^-$-space and a $\lambda^+$-space is
always the zero vector.
On the other hand, the eigenvectors of $T$ corresponding to eigenvalues
not greater than $\lambda$  span a
 $\lambda^-$-space respectively
the eigenvectors corresponding to eigenvalues that greater than $\lambda$
span a $\lambda^+$-space and the sum of their
$\dim_{\cR_i}$-dimensions is $1$.\qed
\begin{lemma}
For any $T\in \cR_i$,
$$\sigma^i_T(\lambda)= \sigma^{i+1}_{\phi_i(T)}(\lambda)\,.$$
$$\ws^i_T(\lambda)= \ws^{i+1}_{\phi_i(T)}(\lambda)\,.$$
\end{lemma}
\proof
First note that $\phi_i$ does not only map $\cR_i$ into
$\cR_{i+1}$ but also it naturally associates
$\cR_{i+1}$-submodules of $V_{i+1}$ to $\cR_{i}$-submodules of  $V_{i}$
preserving their dimensions, that is  for any $\cR_i$-submodule $L$
$$\dim_{\cR_i}(L)=\dim_{\cR_{i+1}}(\phi_i(L))\,.$$
Simply let $P_L$ be the orthogonal projection onto $L$, then we define
$\phi_i(L)$ as the range of $\phi_i(P_L)$.
If $L\subseteq V_i$ is a $\lambda^-$-space of $T$, then
$\phi_i(L)$ is a $\lambda^-$-space of $\phi_i(T)$. Similarly,
if $N\subseteq V_i$ is a $\lambda^+$-space of $T$, then
$\phi_i(N)$ is a $\lambda^+$-space of $\phi_i(T)$. Hence
the image of a maximal dimensional $\lambda^-$-space must
be a maximal dimensional $\lambda^-$-space as well. \qed

\noindent
By the previous lemma, we have spectral functions
$\sigma_T(\lambda)$ and $\ws_T(\lambda)$ on the inductive
limit ring
$\lim_{\stackrel{\phi_i}{\to}} \cR_i$.
\begin{proposition} \label{spec}
Let $\cR$ be the rank closure of
$\lim_{\stackrel{\phi_i}{\to}} \cR_i$ as in Section \ref{sec2}.
Suppose that $A_i\in\cR_i$, $A\in\cR$ and $A_i\to A$ in the metric
defined by the rank function. Then the
spectral functions $\sigma_{A_i}$ uniformly converge to a function
$\sigma_{A}$ and $\sigma_{A}$ does not depend on the choice
of the sequence $\{A_i\}^\infty_{i=1}$. Also, for any $A,B\in \cR; 
\|\sigma_A-\sigma_B\|_\infty \leq d(A,B)$ where
$\|\,\|_\infty$ is the $L^\infty$-norm.
\end{proposition}
\proof
First we need the following comparison lemma.
\begin{lemma} \label{l3}
Let $T,S\in \cR_i$ such that $r_i(T-S)=\epsilon.$
Then for any $\lambda\geq 0$,
\begin{equation}
\label{e14}
|\sigma^i_T(\lambda)-\sigma^i_S(\lambda)|\leq \epsilon
\,\,\mbox{and}\,\,\,
|\ws^i_T(\lambda)-\ws^i_S(\lambda)|\leq \epsilon
\end{equation} \end{lemma}
\proof
Let $L$ be a $\lambda^-$-space for $T$. Note that
$\dim_{\cR_i}\ke (T-S)=1-\epsilon$, since
$r_i(T-S)=\dim_{\cR_i} \im (T-S)$.
Also, $$\dim_{\cR_i} \ke (T-S)+\dim_{\cR_i} L\leq
1+\dim_{\cR_i}(\ke (T-S)\cap L)\,.$$
That is
$$\dim_{\cR_i}(\ke (T-S)\cap L)\geq \dim_{\cR_i} L-\epsilon\,.$$
Clearly, $\ke (T-S)\cap L$ is a $\lambda^-$-space for $S$, thus
our lemma follows. \qed

\noindent
Now we finish the proof of our proposition. By (\ref{e14}),
the spectral function $\sigma: \lim_{\stackrel{\phi_i}{\to}} \cR_i\to
L^\infty[0,\infty)$ is
a Lipschitz-continuous map from our inductive limit ring to the
Banach space of bounded functions. Hence $\sigma$ extends
to the metric completion $\cR$ in a unique continuous way. \qed

\noindent
Later we shall use the following simple
version of Lemma \ref{l3}.
\begin{lemma}
\label{l4}
If $A,B$ are positive
 self-adjoint linear transformations in $\mat_{n\times n}(\bR)$
and
$\frac{\rank(A-B)}{n}\leq \e$, then for any $\lambda\geq 0$:\,
$|N_A(\lambda)-N_B(\lambda)|\leq \epsilon$.
\end{lemma}
\section{Random Schr\"odinger operators on lattices}
Let $G_d$ be the $d$-dimensional standard lattice graph.
For each vertex $p$ of $G_d$ we have independent, identically
distributed random variables $X_{(p)}$ which take the real values
$J=\{c_1,c_2,\dots,c_k\}$ with probabilities $\{p_1,p_2,\dots,p_k\}$,
$p_i >0$. Now let $\Omega=J^{G_d}$ the associated Bernoulli state
space with the product measure. Thus for each $\omega\in\Omega$, we
have an operator $\Delta_\omega$:
$$\Delta_\omega f(x)=\deg f(x)\,\omega(x)-\sum_{(x,y)\in E(G_d)} f(y)\,.$$
\begin{theorem} \label{t1}
For any \Fo -sequence $\{Q_n\}^\infty_{n=1}\subset G_d$ the following holds:
For almost all $\omega\in\Omega$, the spectral distribution
functions $N_{\Delta_{Q_n}}$ converge uniformly
to a function $N_{\Delta}$, where $N_{\Delta_{Q_n}}$
are the normalized spectral distribution functions of the
truncated Laplacians $p_{Q_n}\Delta_{\omega}i_{Q_n}$, where
$p_{Q_n}:L^2(G_d)\to L^2(Q_n)$ is the standard projection operator
and $i_{Q_n}:L^2(Q_n)\to L^2(G_d)$ is its adjoint, the imbedding operator.
The function $N_{\Delta}$ does not depend on $\omega$.
\end{theorem}
\proof
First of all note that by adding a sufficiently large constant to
the numbers $\{c_1,c_2,\dots,c_k\}$ we may suppose that
$\Delta_\omega$ is positive operator for all $\omega\in\Omega$.
The proof shall be given in several steps.

\noindent
\underline{Step $1.$} We construct a Bratteli system as in Section \ref{sec2}.
Consider the cubes $C_i$ in $G_d$ with sidelength $2^i$. That is
$C_i=\{0,1,2,\dots,2^i-1\}^d$. Now let $A_i$ be the finite
set of configurations ($J$-valued functions) on $C_i$.
Then $k_i=|A_i|=k^{2^{id}}$.

\noindent \underline{Step $2.$} Now we construct the weights
$w_{\alpha,\beta}$. For each configuration $a^{i+1}_\beta\in
A_{i+1}$ we associate $2^d$ (not necessarily different) elements of
$A_i$. Namely, the cube $C_{i+1}$ is partitioned into $2^d$ dyadic
cubes of sidelength $2^i$ and the configuration $a^{i+1}_\beta$
determines a configuration on each such subcubes. Then
$w_{\alpha,\beta}$ is the number of occurences of the configuration
$a^i_\alpha\in A_i$ in $a^{i+1}_\beta$.

\noindent
\underline{Step $3.$} Now we define the parameters of our Bratelli
diagrams. First of all $n_{i,\alpha}=2^{id}$ for any $1\leq\alpha\leq
k_i$. Now let $p_{i,\alpha}$ be the probability of the configuration
$a^i_\alpha$. That is
$$p_{i,\alpha}:=\prod_{x\in C_i} P(a^i_\alpha(x))\,,$$ where
of course $P(c_i)=p_i$. Thus, $\sum^{k_i}_{\alpha=1} p_{i,\alpha}=1$
holds.

\noindent
\underline{Step $4.$}
\begin{lemma}
\label{l19}
For any $1\leq\alpha\leq k_i$
$$p_{i,\alpha}=\sum^{k_{i+1}}_{\beta=1} M(\beta,\alpha) p_{i+1,\beta}\,.$$
\end{lemma}
\proof
Le us consider the cube $C_{i+1}$ with sidelength $2^{i+1}$ that determines
$2^d$ dyadic subcubes of sidelength $2^i$. Let us randomly choose
a $J$-valued function on $C_{i+1}$ (the probability of having the value
$c_i$ on the vertex $x$ is $p_i$). The expected value of the number of
configurations of the type $a^i_\alpha$ on the $2^d$-subcubes is
of course $2^d p_{i,\alpha}$. By the theorem of conditional expectations
this value equals to $\sum^{k_{i+1}}_{\beta=1} p_{i+1,\beta} w_{\alpha,\beta}$,
where $w_{\alpha,\beta}$ is the number of occurences of the
configuration $a^i_\alpha$ in $a^{i+1}_\beta$. Since $w_{\alpha,\beta}=
2^d M(\beta,\alpha)$, our Lemma follows. \qed

\noindent
\underline{Step $5.$} The finite dimensional algebra $\cR_i$ is defined
as $\bigoplus^{k_i}_{\alpha=1} \mat_{2^{id}\times 2^{id}}(\bR)$, that is
each element of $\cR_i$ is represented by $k_i$
$2^{id}$ by $ 2^{id}$ matrices. Each configuration
$a^i_\alpha$  defines an element
$M(a^i_\alpha) \in\mat_{2^{id}\times 2^{id}}(\bR)$.
$$M(a^i_\alpha) v_x=\deg(x) P(a^i_\alpha(x))
v(x)-\sum_{y\in C_i \mid (x,y)\in G_d}
v_y\,$$
where $v_x$ is the basis vector of $\bR^{2^{id}}$ associated to
the vertex $x\in C_i$ (for the sake of simplicity we identify matrices
and linear transformations). Note that $\deg(x)$ is the degree of vertex $x$
in the subgraph spanned by $C_i$.
Therefore we defined an element $\widetilde{\Delta}_i\in \cR_i\subset\cR$,
where $\cR$ is the metric completion of the inductive limit ring.

\noindent
\underline{Step $6.$}
\begin{lemma} \label{l21}
The elements $\{ \widetilde{\Delta}_i\}^\infty_{i=1}$ form a \\ Cauchy-sequence
in $\cR$, hence the limit $\limi \widetilde{\Delta}_i=\widetilde{\Delta}$
is well-defined.\end{lemma}
\proof
We need to estimate the rank $r(\widetilde{\Delta}_i-\widetilde{\Delta}_j)$.
First we partition $C_j$ into $2^{d(j-i)}$ dyadic subcubes. Any such
subcube has boundary points and non-boundary points. Boundary points
are those points which are adjacent to a point of another subcube.
By our definition,
$$\widetilde{\Delta}_i(v_x)=\widetilde{\Delta}_j(v_x)\,$$
for any non-boundary point $x\in C_j$. Hence
$r(\widetilde{\Delta}_i-\widetilde{\Delta}_j)$ is bounded above
by $\beta_{i,j}$, where
$$\beta_{i,j}:=\frac{\{\mbox {the number of boundary points in $C_j$.}\}}
{|C_j|}\,.$$
Since for large enough $i$, $\beta_{i,j}<\epsilon$ for any $j>i$, the
lemma follows. \qed

\noindent
\underline{Step $7.$}
Now we turn to the proof of Theorem \ref{t1}. Let $\{Q_n\}^\infty_{n=1}$
be a \Fo -sequence in $G_d$. Let $\omega\in\Omega$ be a configuration
on $G_d$. Pick a positive integer $j$ and cover $G_d$ by disjoint
translates of our standard $C_j$-cube. That is
$G_d=\bigcup_{\underline{t}}\{C_j+\underline{t}\}$, 
where all the coordinates of the vectors
$\underline{t}$ are divisible by $2^j$. We compare the following
two self-adjoint transformations
$$\Delta_{Q_n}:L^2(Q_n)\to L^2(Q_n)\,$$
given by $\Delta_{Q_n}=p_{Q_n}\Delta_\omega i_{Q_n}$, and
$$\Delta^j_{Q_n}:L^2(Q_n)\to L^2(Q_n)\,$$
given by $\sum_{w} p_{Q_n}\Delta_\omega i_{w} $,
where $w$ runs through all the translated copies of $C_i$ which
are contained completely in the set $Q_n$.

\noindent
By the \Fo -property, for any $\epsilon>0$ there exist integers
$j_\e,n_\e$ such that
\begin{itemize} \item
$\frac{\rank(\Delta_{Q_n}-\Delta_{Q_n}^{j_\e})}{|Q_n|}<\e\,.$
\item
and for any $n>n_\e$, $\| s_{\widetilde{\Delta}}-
s_{\widetilde{\Delta}_{j_\e}}\|_\infty<\e $\,
where $\|\,\|_\infty$ denotes the $L^\infty$-norm. \end{itemize}
Hence by Lemma \ref{l4}, if $n>n_\e$
\begin{equation}
\label{e24F}
| N_{\Delta_{Q_n}}(\lambda)-N_{\Delta_{Q_n}^{j_\e}}(\lambda)|<\e\,,
\end{equation}
for any $\lambda\geq 0$.

\noindent
\underline{Step $8.$}
In order to finish the proof of our theorem it is enough to prove
that for any $\e>0$
$$\limsup_{n\to\infty} \|s_{\widetilde{\Delta}_{j_\e}}-
 N_{\Delta^{j_\e}_{Q_n}}\|_\infty\leq \e\,$$
shall hold for almost all $\omega\in\Omega$. The function $N_\Delta$
is defined as $s_{\widetilde{\Delta}}$.

\noindent
Again we consider the set of all configurations $A_{j_\e}$ on the
cube $C_{j_\e}$. Then by Lemma \ref{l1}:
$$s_{\widetilde{\Delta}_{j_\e}}(\lambda)=\sum_{w_\alpha\in A_{j_\e}}
p_{j_\e,\alpha} N_{w_\alpha}(\lambda)\,,$$
where $N_{w_\alpha}$ is the normalized spectral distribution function
of the transformation $\Delta_{w_\alpha}$.
On the other hand,
$$ N_{\Delta^{j_\e}_{Q_n}}(\lambda)=\sum_{w_\alpha\in A_{j_\e}}
R^n_{j_\e,\alpha} N_{w_\alpha}(\lambda)\,,$$
where $R^n_{j_\e,\alpha}$ is the number of occurences of the configuration
$w_\alpha$ in the $C_{j_\e}$-translates contained by $Q_n$ divided
by the number of all such translates.
By the Theorem of Large Numbers, for almost all $\omega\in\Omega$:
$\limn R^n_{j_\e,\alpha}=P_{j_\e,\alpha}$
holds for all configuration $w_\alpha\in A_{j_\e}$. Hence the Theorem
follows. \qed
\section{ Percolation Hamiltonians}
In this section we apply our approximation method to prove uniform
spectral convergence in the case of bond- and site-percolation
Hamiltonians. The pointwise spectral convergence
was proved in this case in \cite{KM} and \cite{V}.

\noindent \underline{The bond percolation Hamiltonian:\,} Again we work
on the lattice $G_d$. For the edges $e\in E(G_d)$ we consider
independent indentically distributed random variables $X_{(e)}$,
such that
$$P(X_{(e)}=0)=p, \quad P(X_{(e)}=1)=1-p\,.$$
The associated Bernoulli-state space $\Omega^B=\prod_{e\in E(G_d)}\{0,1\}$
is equipped with the standard product measure. Then each $\omega\in \Omega$
defines a subgraph $G^B_\omega\subset G_d$, where the vertex set is
the lattice and $e\in E(G^B_\omega)$ if and only if $\omega(e)=0$.

\begin{theorem}
\label{t2} For all \Fo -sequences $\{Q_n\}^\infty_{n=1}\subset G_d$
( in the original sense in the lattice) the following statement
holds: For almost all $\omega\in\Omega$ the normalized spectral
distribution functions $N_{\Delta_{Q_n}}$ uniformly
converge to the integrated density of state function $N^B_\Delta$.
The function $N^B_\Delta$ does not depend on $\omega$. The
normalized spectral distribution functions
$N_{\Delta_{Q_n}}$  are associated to the Laplacian
operator of the subgraphs of $G^B_\omega$ spanned by the vertices of
$Q_n$.
\end{theorem}

\noindent
\underline{The site percolation Hamiltonian:\,}
In the case of site percolation Hamiltonian we consider
independent identically distributed random variables
(with the same distribution as above)
indexed by the vertices of the lattice $G_d$. Again, the state space
$\Omega^S=\prod_{p\in G_d }\{0,1\}$ is equipped with the product measure.
For $\omega\in\Omega^S$ one has the subgraph $G^S_\omega\subset G^d$,
where $e\in E(G^S_\omega)$ if and only if $\omega(p)=\omega(q)=0$ for
the endpoints of $e$.

\begin{theorem}
\label{t3} For all \Fo -sequences $\{Q_n\}^\infty_{n=1}\subset G_d$
the following statement
holds: For almost all $\omega\in\Omega$ the normalized spectral
distribution functions $N_{\Delta_{Q_n}}$ uniformly
converge to the integrated density of state function $N^S_\Delta$.
The function $N^S_\Delta$ does not depend on $\omega$. The
normalized spectral distribution functions
$N_{\Delta_{Q_n}}$ are associated to the Laplacian
operator of the subgraphs of $G^S_\omega$ spanned by the vertices of
$Q_n$.
\end{theorem}
\proof (of Theorem \ref{t2}, the proof of Theorem \ref{t3} is
completely analogous)

\noindent
\underline{Step $1.$} Let $A_i$ be the finite configuration
space of all $\{0,1\}$-valued functions on lattice edges in the
$2^i$-cube $C_i$, $k_i=|A_i|\,.$

\noindent
\underline{Step $2.$} For $a^i_\alpha\in A_i$ and $a^{i+1}_\beta\in A_{i+1}$
let $w_{\alpha,\beta}$ be the number of occurences of the configuration
$a^i_\alpha\in A_i$ in the $2^d$ subconfigurations determined by
$a^{i+1}_\beta\in A_{i+1}$.

\noindent
\underline{Step $3.$} For $a^i_\alpha\in A_i$, let $p_{i,\alpha}$ the
configuration probability that is
$$p_{i,\alpha}:=p^{T(a^i_\alpha)} (1-p)^{S(a^i_\alpha)}\,,$$
where $T(a^i_\alpha)$ is the number of zeroes in the configuration
$a^i_\alpha$ and $S(a^i_\alpha)$ is the number of ones in the configuration
$a^i_\alpha$.

\noindent
\underline{Step $4.$}
$$p_{i,\alpha}=\sum^{k_{i+1}}_{\beta=1} M(\beta,\alpha) p_{i+1,\beta}\,.$$
follows from the conditional expectation argument as in the proof
of Theorem \ref{t1}.

\noindent
\underline{Step $5.$}
 The finite dimensional algebra $\cR_i$ is again isomorphic to
 $\bigoplus^{k_i}_{\alpha=1} \mat_{2^{id}\times 2^{id}}(\bR)$.
 Each configuration
$a^i_\alpha$  defines an element
$M(a^i_\alpha) \in\mat_{2^{id}\times 2^{id}}(\bR)$ as
the associated Laplacian operator $\Delta_{a^i_\alpha}:L^2(C_i)\to L^2(C_i)$.
Hence again we defined an element $\widetilde{\Delta}_i\in \cR_i\subset\cR$.

\noindent
\underline{Step $6.$} The fact that the elements
$\{\widetilde{\Delta}_i\}^\infty_{i=1}$ form a Cauchy sequence
in $\cR$ can be seen exactly the same way as in the proof of Theorem \ref{t1}.

\noindent
\underline{Step $7.$}
 Let $\{Q_n\}^\infty_{n=1}$
be a \Fo -sequence in $G_d$. Let $\omega\in\Omega$ be a configuration
on $G_d$. We consider the same translated copies
as in the proof of Theorem \ref{t1}. Let us compare the following
two self-adjoint transformations:
$$\Delta_{Q_n}:L^2(Q_n)\to L^2(Q_n)\,$$
where $\Delta_{Q_n}$ is the Laplacian of the subgraph of $G^B_\omega$
spanned by the vertices of $Q_n$ and
$$\Delta^j_{Q_n}:L^2(Q_n)\to L^2(Q_n)\,$$
given by $\sum_{w} \Delta_w$,
where $\Delta_w$ is the Laplacian associated to the dyadic subcube $w$.
Again, by the \Fo -property
for any $\epsilon>0$ there exist integers
$j_\e,n_\e$ such that
\begin{itemize} \item
$\frac{\rank(\Delta_{Q_n}-\Delta_{Q_n}^{j_\e})}{|Q_n|}<\e\,.$
\item
and for any $n>n_\e$, $\| s_{\widetilde{\Delta}}-
s_{\widetilde{\Delta}_{j_\e}}\|_\infty<\e $\,. \end{itemize}
Hence by Lemma \ref{l4}, if $n>n_\e$
\begin{equation}
| N_{\Delta_{Q_n}}(\lambda)-N_{\Delta_{Q_n}^{j_\e}}(\lambda)|<\e\,,
\end{equation}
for any $\lambda\geq 0$.

\noindent
\underline{Step $8.$} The final argument, that for any $\epsilon>0$
$$\limsup_{n\to\infty} \|s_{\widetilde{\Delta}_{j_\e}}-
 N_{\Delta^{j_\e}_{Q_n}}\|_\infty\leq \e\,$$
holds for almost all $\omega\in\Omega$ follows exactly the same way
as in the case of the random Schr\"odinger operators. \qed

\section{Pattern-invariant operators on self-similar graphs}

Let $G(V,E)$ be a connected infinite graph with bounded vertex degrees.
A pattern-invariant operator $\cA:L^2(V)\to L^2(V)$
(given by its operator kernel $A:V\times V\to \bR$)
satisfies the following properties:
\begin{itemize}
\item
$A(x,y)=0$ if $d_G(x,y)>r_{\cA}$, where $d_G$ is the shortest path
distance on the vertices. The value $r_{\cA}$, the propagation of $\cA$,
depends only on $\cA$.
\item
$A(x,y)=A(\psi(x),\psi(y))$ if $\psi$ is a graph isomorphism
(there can be more than one) between the $r_{\cA}$-ball around $x$
and the $r_{\cA}$-ball around $\psi(x)$, mapping of course $x$ to
$\psi(x)$.
\end{itemize}
Note that
$$\cA(f)(x)=\sum_{y\in V} A(x,y) f(y)\,,$$ for $f\in L^2(V)$.
These sort of operators were considered in \cite{LS1},\cite{LS2} and \cite{V}.

\noindent
Now we define self-similar graphs (there are plenty of definitions
we choose one which fits our purposes).
 First we fix two positive integers $d$ and $k$.
Let $G_1$ be a finite connected graph with
vertex degree bound $d$ with a distinguished subset
$S_1\subset V_1(G_1)$, which we call the set of connecting vertices.
Now we consider the graph $\widetilde{G}_1$, which consists of $k$
disjoint copies of $G_1$ with following additional properties:
\begin{itemize}
\item The graph $G_1$ is identified with the first copy.

\item In each copy the vertices associated to a connecting vertex
of $G_1$ is a connecting vertex of the graph $\widetilde{G}_1$.
\end{itemize}
The graph $G_2$ is defined by adding some edges to $\widetilde{G}_1$
such that both endpoints of these new edges are connecting vertices.
The resulting graph should still have vertex degree bound $d$.
Finally the subset $S_2\subset V(G_2)$ is chosen as a subset of the
connecting vertices of $\widetilde{G}_1$ such that $S_2\cap
V(G_1)=\emptyset.$ That is $G_1\subset G_2$ is a subgraph and
vertices of $G_2$ which are not in $G_1$ are connecting vertices.
Inductively, suppose that the finite graphs $G_1\subset G_2\subset
\dots \subset G_n$ are already defined and the vertex degrees in
$G_n$ are not greater than $d$. Also suppose that a set $S_n\subset
V(G_n)$ is given and $S_n\cap V(G_{n-1})=\emptyset$. Now the graph
$\widetilde{G}_n$ consists of $k$ disjoint copies of $G_n$ and
\begin{itemize}
\item The graph $G_n$ is identified with the first copy.

\item In each copy the vertices associated to a connecting vertex
of $G_n$ is a connecting vertex of the graph $\widetilde{G}_n$.
\end{itemize}

Again, $G_{n+1}$ is constructed by adding edges to $\widetilde{G}_n$
with endpoints which are connecting vertices, preserving the vertex
degree bound condition. The set of connecting vertices $S_{n+1}$ is
chosen as a subset of the connecting vertices of $\widetilde{G}_n$,
such that $S_{n+1}\cap V(G_n)=\emptyset\,.$ The union of the graphs
$\{G_n\}^\infty_{n=1}$ is connected infinite graph with vertex
degrees not greater than $d$. The graph $G$ is {\it self-similar} if
$\limn \frac{|S_n|}{|V(G_n)|}=0$. In this case
$\{V(G_n)\}^\infty_{n=1}$ form a \Fo -sequence of $G$. Note that all
Euclidean lattices is self-similar in our sense.
\begin{theorem}
\label{t4}
Let $G$ be a self-similar graph and $\cA$ be a self-adjoint pattern-
invariant operator. Then there exists a real function
$N_{\cA}$ such that for any \Fo -sequence $\{Q_n\}^\infty_{n=1}\subset V(G)$,
$N_{\cA_{Q_n}}$ uniformly converge to $N_{\cA}$, where
$N_{\cA_{Q_n}}$ are the spectral distribution functions of the
truncated operators $p_{Q_n}\cA\, i_{Q_n}$.
\end{theorem}
In this case we have the simplest possible Bratteli system (the one
originally considered by John von Neumann):
$$\mat_{l\times l}(\bR)\stackrel{\phi_1}{\to}
\mat_{kl\times kl}(\bR)\stackrel{\phi_2}{\to} \dots $$
That is $\cR_i=\mat_{k^{i-1}l\times k^{i-1}l}(\bR)$, where $l=|V(G_1)|$.
For each $i\geq 1$, $k_i=1$ and each weight equals to $k$.
Then $\widetilde{\Delta}_i\in\cR_i$ is just the operator $\cA_{G_i}$.
The proof of Theorem \ref{t4} is completely analogous
 to the one of Theorem \ref{t1}
and left for the interested reader.
\section{A general conjecture}\label{last}
One can easily see that for our self-similar graphs, uniform spectral
convergence holds even for the random Schr\"odinger operators
and percolation Hamiltonians. For nice Markovian substitution
systems our method works without any trouble. The most general graphs
for which universal spectral convergence might hold are the following
ones.

\noindent
Let $G(V,E)$ be a connected infinite graph with bounded vertex degrees.
We suppose that $G$ is an amenable graph, that is $G$ has \Fo -sequence
$\{F_n\}^\infty_{n=1}$. The $r$-pattern of a vertex $x\in V$ is
the graph automorphism class of the rooted ball around $x$.
That is $x,y\in V$ have the same $r$-pattern if there exists a
graph isomorphism $\phi$ between the balls $B_r(x)$ and $B_r(y)$ such
that $\phi(x)=y$. Denote by $P_r(G)$ the finite set of all possible
$r$-patterns in $G$. We say that $G$ is an {\it abstract quasicrystal graph}
if for any $\alpha\in P_r(G)$ there exists a {\it frequency} $P(\alpha)$
such that
\begin{itemize}
\item
For {\it any} \Fo -sequence $\{Q_n\}^\infty_{n=1}$:\,
$$\limn \frac{|Q^\alpha_n|}{|Q_n|}=P(\alpha)\,,$$
where $Q^\alpha_n\subseteq Q_n$ is the set of vertices in $Q_n$ having
the $r$-pattern $\alpha$. \end{itemize}

\noindent
The graphs considered by Lenz and Stollmann are such abstract quasicrystal
graphs.
Another interesting examples are the Cayley graphs of finitely
generated amenable groups.
\begin{conjecture}
If $G$ is an abstract quasicrystal graph, then for the random Schr\"odinger
operators, percolation Laplacians as well as self-adjoint pattern invariant
operators associated to $G$ the uniform spectral convergence exists.
\end{conjecture}
Note that for the Laplacian operator the existence of the integrated
density of states in the weak sense of convergence follows from the
argument of Theorem 3. \cite{Elek}. Even for the general pattern-invariant
operators and the random operators the limit graph method yields
the existence of the integrated density of states. The problem is
to handle the jumps of the integrated density of states. One should
prove that for each value $\lambda$, where
the integrated density of state function jumps
the normalized dimensions of the eigenfunctions of the truncated
operators on the \Fo -sequence converge exactly to the size of the jump.
The same problem emerges in the theory of $L^2$-invariants in the
so-called Approximation Theorem (see \cite{Lueck}).

\end{document}